# First-principles prediction of pressure dependent mechanical, electronic, optical, and superconducting state properties of *NaC$_6$*: A potential high-*T*$_c$ superconductor

Nazmun Sadat Khan[1], B. Rahman Rano[1], Ishtiaque M. Syed[1], R. S. Islam[2], S. H. Naqib[2,*]

[1]Department of Physics, University of Dhaka, Dhaka-1000, Bangladesh
[2]Department of Physics, University of Rajshahi, Rajshahi-6205, Bangladesh
*Corresponding author e-mail: salehnaqib@yahoo.com

**Abstract**

Very recently carbon-rich *NaC$_6$* with sodalite-like structure has been predicted to show high superconducting transition temperature above 100 K at relatively low applied (compared to high-*T*$_c$ hydrides) hydrostatic pressures. We have investigated the pressure dependent structural, elastic, electronic, superconducting state, and optoelectronic properties of *NaC$_6$* in this study. Some important thermophysical properties have also been explored. The elastic properties along with Poisson's and Pugh's ratios and optoelectronic parameters are investigated for the first time. *NaC$_6$* was found to be structurally stable only at high pressures at and above 40 GPa, in agreement with previous study. The compound is highly ductile and the chemical bonding is prominently metallic in nature. The Debye temperature shows strong pressure dependence. The Grüneisen parameter also exhibits significant pressure dependence. The electronic band structure reveals metallic character and consists of highly dispersive and almost flat bands crossing the Fermi level. Both the electronic density of states at the Fermi level and repulsive Coulomb pseudopotential increase gradually with increasing pressure in the range 40 GPa to 70 GPa. The degree of dispersion in the *E*(*k*) curves depend weakly on pressure both in the valence and conduction bands. The optical parameters spectra, studied for the first time, correspond well with the electronic band structure. *NaC$_6$* absorbs and reflects electromagnetic radiation quite efficiently in the mid-ultraviolet region. Superconducting transition temperatures of *NaC$_6$* have been estimated at different pressures and compared with previously reported values. The effects of various parameters on *T*$_c$ have been discussed in details.

## 1. Introduction

The discovery of superconductors with high transition temperatures has attracted significant amount of research interest, both for its myriads of potential applications and the theoretical significance of the phenomenon [1–3]. However, a major stumbling block to its applications is the difficulty of achieving the conditions in which such a superconducting state can exist, namely requiring both high pressures and low temperatures. This is one of the primary reasons why there exists great interest in finding high-*T*$_c$ superconductors under external conditions conducive for practical applications. In this paper, high-$T_c$ superconductors refer to those compounds other than the cuprates whose superconducting critical temperatures are at or above the boiling point of liquid nitrogen, meaning it could be cooled by relatively inexpensive coolants, as opposed to room temperature superconductors which have $T_c$ at 0°C or above.

One of the earliest breakthroughs in this search was the theoretical prediction of high-$T_c$ superconductivity of metallic hydrogen at very high pressures [4,5]. As of now, this superconductivity in metallic hydrogen has not been demonstrated, but it nevertheless inspired searches for high temperature superconductivity in hydrogen-rich compounds. A number of candidates have emerged such as $H_3S$, $H_2S$, $CaH_6$, $LaH_{10}$ and $YH_{10}$ [6–8]. Recently, room-temperature superconductivity has been reported in carbonaceous sulfur hydride ($CH_8S$) [9]. Of the materials mentioned, $YH_6$ and $YH_{10}$ are notable, because the hydrogen in those compounds form a sodalite-like structure [8]. Neil Ashcroft predicted on theoretical grounds that hydrogen under high external pressure will exhibit superconductivity, however the need for such external physical pressure might be alleviated by forming metallic hydrides where the hydrogen atoms are "pre-compressed" because of its chemical structure [4,10]. Prior research indicates this to be true for the case of $CaH_6$, $YH_6$ and $YH_{10}$ having sodalite-like hydrogen cage [7,11]. The nature of superconductivity in these materials is believed to be conventional, arising out of large electron-phonon coupling and very high Debye temperature.

Despite these advancements, the pressure requirements of these materials for superconductivity to manifest are still quite high. An alternative approach would be to replace $H$ with $C$ or other light elements in those sodalite structures. Superconductivity has been shown to occur in carbon compounds previously, such as boron-doped diamond, and of special significance to our work, in carbon compounds with similar sodalite structures [12–14]. From first-principles calculations, it has been shown that $NaC_6$ exhibits high critical temperature at around 100 K at relatively low pressures [12,13,15]. High electronic density of states and the presence of degenerate states near the Fermi level are significant contributors to a large electron-phonon interaction constant $\lambda$. The characteristic phonon energy is also high due to light mass of C atoms and high stiffness of the crystal lattice. All these contribute to the high $T_c$ of $NaC_6$. The material is stable at (comparatively) modest pressures which is desirable in a candidate for practical high-$T_c$ superconductors.

In this work, we have conducted a thorough first-principles investigation of the physical properties of $NaC_6$ limited not only to its significance as a potential high-$T_c$ superconductor, but also other properties of the material which have not been studied in any great depth till date. We have studied the bulk elastic properties of the material, Debye temperature, Grüneisen parameter, as well as its optical properties. The band structure and energy density of states have been analyzed. We have paid attention to the effect of pressure on the mechanical, electronic, and optical properties of this material. Potential applications of $NaC_6$ depend significantly on understanding these pressure dependent physical characteristics. As far as superconductivity itself is concerned, we have found that the large electron phonon coupling constant and high Debye temperature play a large part in the high critical temperature of $NaC_6$ in fair agreement with the previous reports [12,13]. A detailed analysis of the pressure dependence of $T_c$ has been carried out.

We have organized our manuscript as follows: In Section 2, we state the computation methodology used for all first principles calculations. In Section 3, we have tabulated and analyzed the results of the computations. We have discussed, in order, the crystal structure, the elastic properties, the Debye temperature, electronic band structure and energy density of states, optical properties and the critical temperature of the material under different pressures. In Section 4, we have summarized the key features and principal findings of our study.

## 2. Computational methodology

We have calculated the properties of $NaC_6$ from first principles calculations through CAmbridge Serial Total Energy Package (CASTEP) [16]. Density Functional Theory (DFT) formalism is the basis on which the various properties of $NaC_6$ were derived. In DFT, instead of solving the many-body Schrödinger equation directly where the system is characterized by the complicated and difficult-to-compute electron wave functions, the problem is restated in terms of equivalent non-interacting problem where the system is described in terms of electron density. Solving the Kohn-Sham equations leads to the ground state density of the crystal system [17]. Both Local Density Approximation (LDA) and Generalized Gradient Approximation (GGA-PBE) were used as the exchange correlation functionals [18,19]. It is known that LDA tends to underestimate lattice parameters, while GGA sometimes overestimates those [16]. Ultra-soft pseudopotential, introduced by Vanderbilt, was used during the geometry optimization as it allows for lower basis cut-off, without sacrificing transferability too much [16,20]. Broyden–Fletcher–Goldfarb–Shanno (BFGS) algorithm was used to optimize the geometry of the crystal [21]. Pseudo atomic calculations were performed using the following electronic orbitals: C [$2s^2\ 2p^6$] and Na [$2s^2\ 2p^6\ 3s^1$]. The plane wave basis set cut-off energy has been set to 410 eV. A 6×6×6 $k$-points mesh has been used based on Monkhorst-Pack scheme [22] for sampling the Brillouin zones (BZs) of $NaC_6$. The system geometry was optimized under an energy tolerance of $5.0 \times 10^{-6}$ eV/atom, maximum force tolerance of 0.01 eV/Å, maximum stress tolerance of 0.02 GPa and maximum displacement tolerance of $5.0 \times 10^{-4}$ Å.

The single crystal elastic constants $C_{ij}$, bulk modulus $B$, shear modulus $G$ were calculated from the *stress-strain* method contained in the CASTEP code. The electron energy dispersion and electronic energy density of states have been obtained from the optimized geometry of $NaC_6$. The optical constants spectra were extracted from the estimated complex dielectric function, $\varepsilon(\omega) = \varepsilon_1(\omega) + i\varepsilon_2(\omega)$, which describes the frequency/energy dependent interactions of photons with the electrons in solids. Using the Kramers-Kronig transformations, the real part $\varepsilon_1(\omega)$ of dielectric function $\varepsilon(\omega)$ was calculated from the imaginary part $\varepsilon_2(\omega)$. The imaginary part, $\varepsilon_2(\omega)$, was calculated from the matrix elements of optical transitions between occupied and unoccupied electronic orbitals employing the CASTEP supported formula expressed as:

$$\varepsilon_2(\omega) = \frac{2e^2\pi}{\Omega\varepsilon_0} \sum_{k,v,c} |\langle \psi_k^c | \hat{u}.\vec{r} | \psi_k^v \rangle|^2 \, \delta(E_k^c - E_k^v - E) \tag{1}$$

where, $\Omega$ is the volume of the unit cell, $\omega$ is angular frequency of the incident photon, $e$ is the electronic charge, $\psi_k^c$ and $\psi_k^v$ are the wave functions of electrons in the conduction and valence band with a wave-vector $k$, respectively. The conditions of conservation of energy and momentum during the optical transition are ensured by the delta function in Eqn. 1. Once the real and imaginary parts of the dielectric function $\varepsilon(\omega)$ is known, all the other optical parameters such as refractive index, optical conductivity, reflectivity, absorption coefficient, and energy loss function can be computed from those [23]. The Debye temperature and other thermophysical parameters have been calculated from the elastic constants and elastic moduli of $NaC_6$. All the properties mentioned here have been calculated in the pressure range from 0 GPa to 70 GPa with different pressure intervals.

## 3. Results and analysis

### 3.1. Crystal structure

The crystal structure of $NaC_6$ is cubic, belonging to the space-group *Im-3m*(No-229)*,* as shown schematically in Fig. 1. The $Na$ atoms are embedded in a *BCC* lattice, surrounded by a characteristic cage (sodalite structure) of carbon atoms linked together through $sp^3$ bonds [12]. Each sodium atom is surrounded by 24 carbon atoms. The lattice parameters optimized under both GGA and LDA in the ground state are listed in Table 1. It is seen that the lattice parameters obtained by LDA are lower than those obtained using GGA, which is to be expected. The structural parameters obtained using GGA agree very well to those obtained in the previous study [12]. Since geometry optimization is a vital component in determining the reliability of *ab-initio* calculations we will focus mainly on results obtained via GGA in the sections to follow.

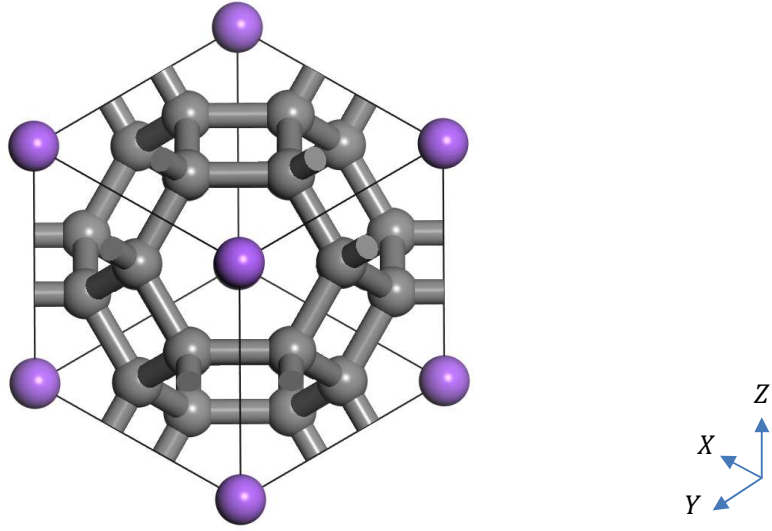

**FIG. 1.** Crystal structure of $NaC_6$. The purple spheres represent the $Na$ atoms at the corners and center of the cubic unit cell, and the grey spheres represent the $C$ atoms forming the sodalite cage structure.

**TABLE 1:** The optimized lattice parameters and cell volumes of $NaC_6$ at 0 GPA obtained from GGA and LDA calculations. The lattice parameters *a*, *b*, and *c* are in Å and the unit cell volume (V) is in Å$^3$.

| Compound (functional) | a | b | c | V |
|---|---|---|---|---|
| $NaC_6$(GGA) | 4.563 | 4.563 | 4.563 | 95.026 |
| $NaC_6$(LDA) | 4.499 | 4.499 | 4.499 | 91.092 |
| $NaC_6$ (Theoretical [12]) | 4.570 | 4.570 | 4.570 | 95.443 |

*3.2. Elastic properties*

$NaC_6$ is predicted to have a high superconducting critical temperature for external pressures at and above 40 GPa. It is therefore, quite important to study the elastic and mechanical properties of the crystal to understand its behavior under high level of mechanical stress. The question of mechanical stability is also important. Because of the high degree of symmetry of cubic crystals, they only have three independent single crystal elastic constants namely, $C_{11}$, $C_{12}$ and $C_{44}$. Stability of the crystal can be investigated via the Born criteria, as modified for cubic crystals [24,25]:

$$C_{11} - C_{12} > 0 \,;\, C_{11} + 2C_{12} > 0 \,;\, C_{44} > 0 \qquad (1)$$

These are necessary and sufficient conditions of crystal stability under static stress. The obtained elastic constants at different values of hydrostatic pressure are given in Table 2.

**TABLE 2:** The single crystal elastic constants for $NaC_6$ ($C_{ij}$ in GPa) calculated at various pressures (GPa).

| Pressure (P) | $C_{11}$ | $C_{12}$ | $C_{44}$ |
|---|---|---|---|
| 0 | 783.20 | 69.86 | -105.63 |
| 5 | 811.67 | 83.06 | -121.56 |
| 10 | 838.29 | 96.34 | -131.80 |
| 15 | 864.39 | 110.16 | -134.44 |
| 20 | 890.21 | 124.62 | -127.46 |
| 25 | 914.65 | 138.44 | -103.86 |
| 29 | 934.03 | 150.01 | -79.51 |
| 30 | 938.60 | 152.96 | -73.51 |
| 31 | 943.78 | 155.64 | -62.78 |
| 32 | 948.11 | 158.87 | -58.11 |
| 33 | 952.73 | 161.85 | -49.28 |
| 35 | 962.56 | 167.19 | -31.39 |
| 40 | 985.68 | 181.95 | 7.80 |
| 50 | 1030.21 | 210.18 | 69.02 |
| 60 | 1073.93 | 238.01 | 102.62 |
| 70 | 1112.59 | 265.94 | 119.84 |

It is noted that the first two Born criteria are satisfied for all the pressures tested. However, the condition $C_{44} > 0$ is satisfied only for pressures close to 40 GPa and above. This indicates that $NaC_6$ is mechanically stable at and above 40 GPa. The phase stability of $NaC_6$ has been checked in the high pressure regions in Ref. [13]. A noticeable feature of the elastic constants is the extremely high values of the elastic constant $C_{11}$ compared to the rest. The high value of elastic constant $C_{11}$ ($C_{11} = C_{22} = C_{33}$) indicates strong resistance to tensile stress along the *a, b,* and *c* axes. This is also a clear signature of extreme anisotropy in bonding strength present in $NaC_6$. The elastic constant $C_{44}$ is a measure of resistance to shearing strain of the crystal. $C_{44}$ ($C_{44} = C_{55} = C_{66}$) is significantly lower than $C_{11}$ at all pressures, meaning that the material is far less resistant to shearing strain. The mechanical failure mode of $NaC_6$ is thus predicted to be controlled by shearing strain. The resistance of the crystal against volume conserving orthogonal distortion is determined by the constants $C_{12}, C_{13}, C_{23}$. In our case, the constant $C_{12}$ ($C_{12} = C_{13} = C_{23}$) is lower as well, meaning the material displays weak resistance to orthogonal distortion. The Cauchy pressure, given by ($C_{12} - C_{44}$), is positive and high at all pressures, indicating the ductile nature [26] of $NaC_6$.

The elastic moduli for bulk (polycrystalline) systems can be obtained from the single crystal elastic constants. These are listed in Table 3. The elastic moduli for bulk systems were calculated using the Voigt and Reuss approximations (indicated using subscripts V and R, respectively) [27,28]. These approximations provide upper and lower bounds of the parameters, respectively [29]. Hills approximations are the arithmetic mean of the respective Voigt and Reuss approximations, and, thus, provide a middle-ground between the two [30].

Besides the elastic moduli, Table 3 contains other important elastic indicators; the Poisson's ratio and the Pugh's ratio calculated at different pressures.

**TABLE 3:** The bulk moduli of the polycrystalline aggregates $B_R$, $B_V$, $B_H$ (in GPa) and shear moduli $G_R$, $G_V$, $G_H$ (in GPa), Young's moduli ($E$), Poisson's ratio ($v$), and Pugh's ratio ($B_V/G_V$) under different pressures for $NaC_6$. The subscripts R, V and H stand for Reuss, Voigt and Hills approximations, respectively.

| Pressure (GPa) | $B_R$ | $B_V$ | $B_H$ | $G_R$ | $G_V$ | $G_H$ | $E$ | $v$ | Pugh's ratio |
|---|---|---|---|---|---|---|---|---|---|
| 31 | 418.28 | 418.35 | 418.32 | -117.06 | 119.96 | 1.44 | 4.33 | 0.498 | 289.17 |
| 32 | 421.94 | 421.95 | 421.94 | -107.39 | 122.98 | 7.79 | 23.23 | 0.490 | 54.13 |
| 33 | 425.44 | 425.48 | 425.46 | -89.59 | 128.60 | 19.50 | 57.63 | 0.477 | 21.81 |
| 35 | 432.33 | 432.32 | 432.33 | -55.22 | 140.23 | 42.50 | 123.47 | 0.452 | 10.17 |
| 40 | 449.84 | 449.86 | 449.85 | 12.84 | 165.43 | 89.13 | 250.84 | 0.407 | 5.04 |
| 50 | 483.51 | 483.52 | 483.52 | 103.42 | 205.41 | 154.42 | 418.68 | 0.355 | 3.13 |
| 60 | 516.63 | 516.65 | 516.64 | 146.98 | 228.75 | 187.87 | 502.68 | 0.337 | 2.75 |
| 70 | 548.15 | 548.16 | 548.15 | 168.02 | 241.23 | 204.63 | 545.96 | 0.334 | 2.67 |

For mechanically stable state under pressures of 40 GPa and above, the bulk modulus of $NaC_6$ is quite high and the sheer modulus is comparatively lower, indicating the material is more resistant to compression than shape deformation. With increasing pressure, both the bulk moduli and sheer moduli increases, the latter increasing quite sharply. The value of Poisson's ratio $v$ is quite high, close to the mathematical upper limit of 0.50, thus the material is highly incompressible [31,32]. With increasing pressure $v$ decreases, however for up to 70 GPa, it remains above 0.25, which indicates that the material is ductile in nature, supporting the analysis from the Cauchy pressures [33]. Large value of Poisson's ratio also implies that covalent bondings do not contribute significantly in the formation of $NaC_6$[31,32]. Ductility, on the other hand, indicates that ionic/metallic bondings are prominent in this compound. Pugh's ratio [34] is a widely employed parameter to distinguish between brittle and ductile behaviors of solids. Solids with a Pugh's ratio of 1.75 or above exhibit ductility; solids possessing lower Pugh's ratio are expected to show brittleness [34-36]. The unrealistically high values of Pugh's ratio seen in Table 3 at pressures below 40 GPa stem from the fact that $NaC_6$ is elastically unstable for P < 40 GPa. The Young's modulus is a measure of the tensile strength of the material, which for an isotropic material can be calculated from bulk modulus and Poisson's ratio. From the table we see that the value of the Young's moduli increases with pressure, which can be explained by the increase of bulk modulus and decrease of Poisson's modulus with pressure. It is instructive to note that almost all the elastic parameters disclosed in Table 3 exhibit an abrupt change at pressure around 40 GPa. This is indicative of structural instability of the system at pressures below 40 GPa. It should be stressed that the results disclosed in Table 3 are novel and no prior estimates exist to compare.

## 3.3. Debye temperature and Grüneisen parameter

From the computed elastic moduli of the polycrystalline aggregates of $NaC_6$, the Debye temperature $\theta_D$ can be calculated, at various pressures [37]. The Debye temperature is related to a number of important properties of the crystal and often arises in relation to specific heat, melting temperatures and phonon calculations. Furthermore, it is related to electron-phonon coupling constant and the critical temperature of superconductors. According to the Debye model, the Debye frequency $\omega_D$ is the highest allowed phonon frequency in crystal. The Debye temperature is related to $\omega_D$ by the formula $\theta_D = \hbar\omega_D/k_B$. The Debye temperature can be calculated from the average sound velocities as follows [37]:

$$\theta_D = \frac{h}{k_B}\left[\left(\frac{3n}{4\pi}\right)\frac{N_A\rho}{M}\right]^{\frac{1}{3}} v_m \qquad (2)$$

where $h$ is the Planck's constant, $k_B$ is the Boltzmann's constant, $n$ is the number of atoms in the molecule, $N_A$ is the Avogadro's constant, $\rho$ is the density, $M$ is the molecular weight. The average sound velocity $v_m$ is calculated by,

$$v_m = \left[\frac{1}{3}\left(\frac{2}{v_t^3} + \frac{1}{v_l^3}\right)\right]^{-\frac{1}{3}} \qquad (3)$$

where $v_t$ and $v_l$ are the transverse and longitudinal elastic wave velocities, respectively. $v_t$ and $v_l$ can be further calculated from the polycrystalline elastic moduli as follows:

$$v_t = \left(\frac{G}{\rho}\right)^{\frac{1}{2}} \qquad (4)$$

and

$$v_l = \left(\frac{B + \frac{4G}{3}}{\rho}\right)^{\frac{1}{2}} \qquad (5)$$

The calculated values of $\rho$, $v_l$, $v_t$, $v_m$ and Debye temperature $\theta_D$ are listed in Table 4. We see that the values of the Debye temperature are quite high and increases with increasing pressure. This contributes significantly to the high superconducting critical temperature of $NaC_6$. The high values of $\theta_D$ is also an indication of the hardness of the material which agrees well with the elastic moduli from previous section.

**TABLE 4:** The density $\rho$ in g/cm$^3$, longitudinal ($v_l$), transverse ($v_t$) and average ($v_m$) elastic wave velocity in m/s, Debye temperature ($\theta_D$) in K and the Grüneisen parameter ($\gamma$) for different pressures (in GPa), for the polycrystalline aggregate of $NaC_6$.

| Pressure | $\rho$ | $v_l$ | $v_t$ | $v_m$ | $\theta_D$ | $\gamma$ |
|---|---|---|---|---|---|---|
| 40 | 1.92 | 17174.97 | 6799.63 | 7704.76 | 1010.55 | 2.71 |
| 50 | 1.96 | 18708.53 | 8854.24 | 9962.58 | 1316.05 | 2.18 |
| 60 | 2.00 | 19539.28 | 9669.45 | 10853.80 | 1443.34 | 2.03 |
| 70 | 2.04 | 20024.77 | 9997.31 | 11216.11 | 1500.88 | 2.01 |

The crystal density of $NaC_6$ is very low. The Debye temperature and sound velocities increase rapidly with increase in the external pressure.

In addition to Debye temperature, we have calculated the Grüneisen parameter, $\gamma$, for $NaC_6$ at different pressures and presented those in Table 6. Grüneisen parameter is an important thermophysical indicator that gives a measure of lattice anharmonicity. A large value of $\gamma$ implies high level of anharmonicity. The Grüneisen parameter was calculated from the computed values of Poisson's ratio using the following expression [38]:

$$\gamma = \frac{3(1+\nu)}{2(2-3\nu)}$$

It is seen from Table that $\gamma$ is quite large at 40 GPa. Further increase in pressure reduces the Grüneisen parameter.

*3.4. Electronic band structure and energy density of states*

The electronic band structure reveals a wealth of information about any material. The band structure plot (Fig. 2) shows the energies of the bands created by the overlap of energy levels of every atom in the crystal lattice at 40 GPa, along a path passing through certain high symmetry points of the Brillouin zone. The range of energies of the bands shown is from -10 eV to 10 eV, which contains the top of the valence bands and the bottom of the conduction bands. The Fermi level has been set at 0 eV. A fairly dispersive band of band width ~ 4.0 eV crosses the Fermi level along the path defined by X-R. This particular band exhibits hole-like dispersion. There are two more flat bands with electronic character running along M-Γ and Γ-R directions which also cross the Fermi level. The electrons belonging to these bands should have very large effective masses. The overall band structure of $NaC_6$ reveals clear metallic signatures at 40 GPa. The bands along R-M are less dispersive compared to the bands along M-Γ and Γ-R, that is to say they become more dispersive as we move towards the center of the Brillouin zone. With increasing pressure, the bands become more dispersive. At the high symmetry points Γ and R, we find a large number of degenerate electronic states. At the Γ point, some of these degenerate states reach and cross the Fermi Level, $E_F$. These degenerate states explain the high DOS around the Fermi level $N(E_F)$ found in the density of states calculations.

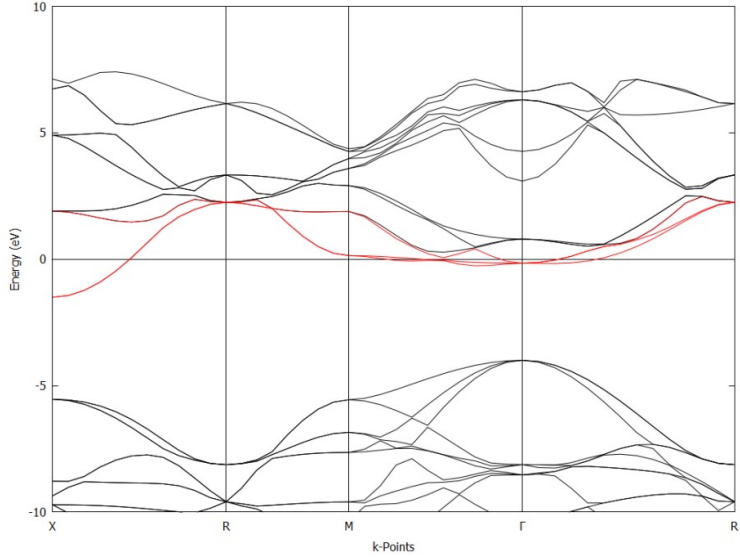

**FIG. 2.** The electronic band structure of $NaC_6$ along points of high symmetry (at 40 GPa) in the first Brillouin zone.

From the total density of states (TDOS) plot derived from the band structure of $NaC_6$ at 40 GPa shown in Fig. 3, it is clear that the density of states near the Fermi level, indicated by the vertical line at 0 eV, is quite high. This high value originates mainly from the nearly flat bands crossing the Fermi level. From the partial density of states (PDOS) calculated for the electronic orbitals of individual atoms, it is seen that most of the contribution towards TDOS comes from the C atoms, while Na atoms have almost no contribution below the Fermi level, and contribute very little at and above the Fermi level, which raises the TDOS slightly. From the DOS graph, the PDOS for C-2s and C-2p are comparable at the lower energy range, indicating that these electronic orbitals hybridize to form $sp^3$ covalent bonds. There is a presence of a shallow pseudogap in the neighborhood of $E_F$. In conventional metallic systems, the pseudogap often refers to the gap or valley between the bonding and anti-bonding peaks. This can occur for a number of reasons like charge transfer, covalence or d-resonance [39]. It should be stressed here that the weak pseudogap feature observed here is fundamentally different from that observed in hole doped high-$T_c$ cuprate superconductors [40-42]. It is also worth noting that for materials having sodalite-like structures, charge transfer between constituent elements and carbon cage is an important contributor towards boosting $T_c$. The Na atoms donate almost all valence electrons to carbon. The presence of the pseudogap around the Fermi level has deep consequences for the nature of the material, such as its electronic stability [43,44]. Since $E_F$ is located towards the right of the pseudogap, close to the antibonding region, $NaC_6$ displays a tendency to be in a disordered state. Figs. 2 & 3 only display the band structure and DOS features of $NaC_6$ at 40 GPa. We have calculated all these properties considering uniform hydrostatic pressures of 50 GPa, 60 GPa, and 70 GPa. At these pressures the compound under consideration is predicted to be structurally stable. Pressure dependent changes in the electronic band structure are monotonic and gradual, characterized by an increase in the electronic energy density of states at the Fermi level.

The TDOS at $E_F$, $N(E_F)$, is a vital parameter for superconductivity. It determines the electron-boson coupling strength [45,46], superconducting condensation energy [47,48], and the magnitude of the repulsive Coulomb pseudopotential, $\mu^*$. This repulsive Coulomb term hinders the formation of Cooper pairs essential for superconductivity [45,46,49].

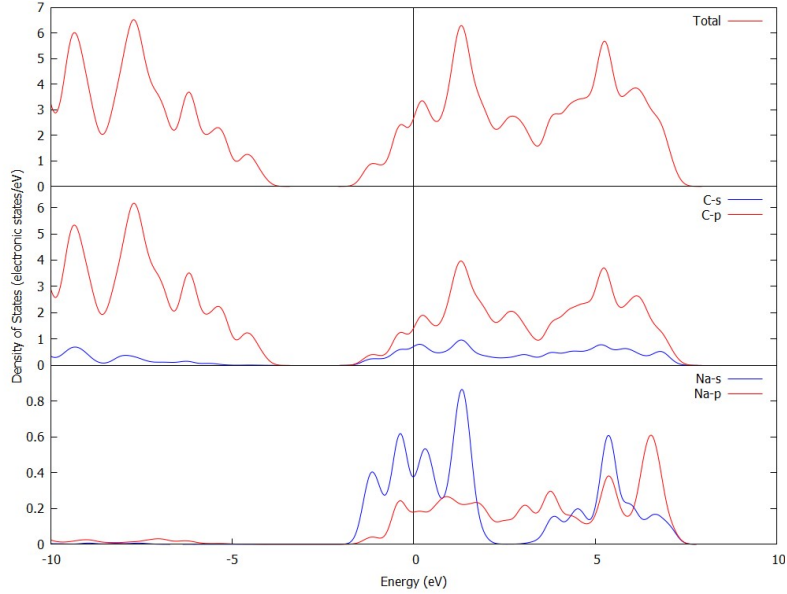

**FIG. 3.** The total and partial density of states of $NaC_6$ for 40 GPa external pressure.

We have summarized the effect of pressure on $N(E_F)$ and $\mu^*$ in Fig. 4. The Coulomb pseudopotentials at different pressures have been calculated following Refs. [50,51]. It is observed that both these parameters increase with increasing pressure. Compared to many other phonon-mediated superconductors, $\mu^*$ for $NaC_6$ is high [45].

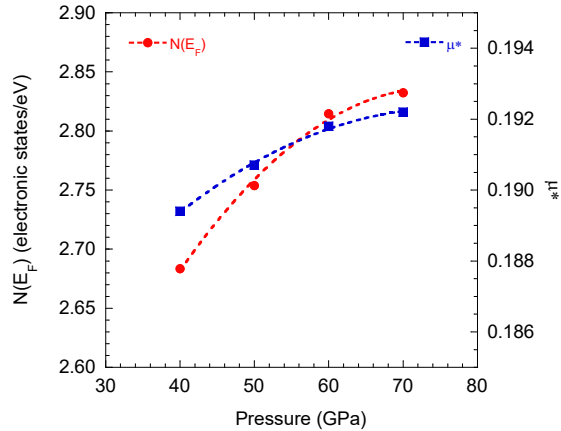

**FIG. 4.** The electronic density of states and repulsive Coulomb pseudopotentials of $NaC_6$ at different pressures. The dashed curves are second-order polynomial fits to the computed values.

Overall, the band structure features seen in this section agree quite well with those found in the previous studies [12,13].

*3.5. Optical properties*

A thorough study of optical parameters as a function of photon energy is important as these parameters describe the behavior of a material in response to incident electromagnetic radiation and are useful in determining the utility of the material in different optoelectronic applications. Besides, the optical parameters spectra are intimately related to the electronic band structure and their study yields complementary information related to fundamental physics. The optical parameters, namely, (a) the complex dielectric function, $\varepsilon(\omega)$ (the real and imaginary parts), (b) refractive index $\eta(\omega)$, (c) optical conductivity $\sigma(\omega)$, (d) reflectivity $R(\omega)$, (e) absorption coefficient $\alpha(\omega)$ and (f) loss function $L(\omega)$ of $NaC_6$ have been calculated at different pressures. All these parameters were obtained by first determining the complex dielectric function. The CASTEP evaluates the imaginary part of the dielectric function using Eqn. 1 [52]. The real part of the dielectric function can then be found using Kramers-Kronig transform. From the dielectric functions, all other optical parameters can be obtained using standard formalism [53]. We have calculated optical parameters for all the pressures mentioned in Section 2. However, we have only disclosed the behaviors of energy dependent optical parameters at two representative pressures, 40 GPa and 60 GPa, in this section. Overall, the effect of pressure is moderate on the optical parameters. This is consistent with the computed electronic band structures where the effect of pressure was also found to be moderate.

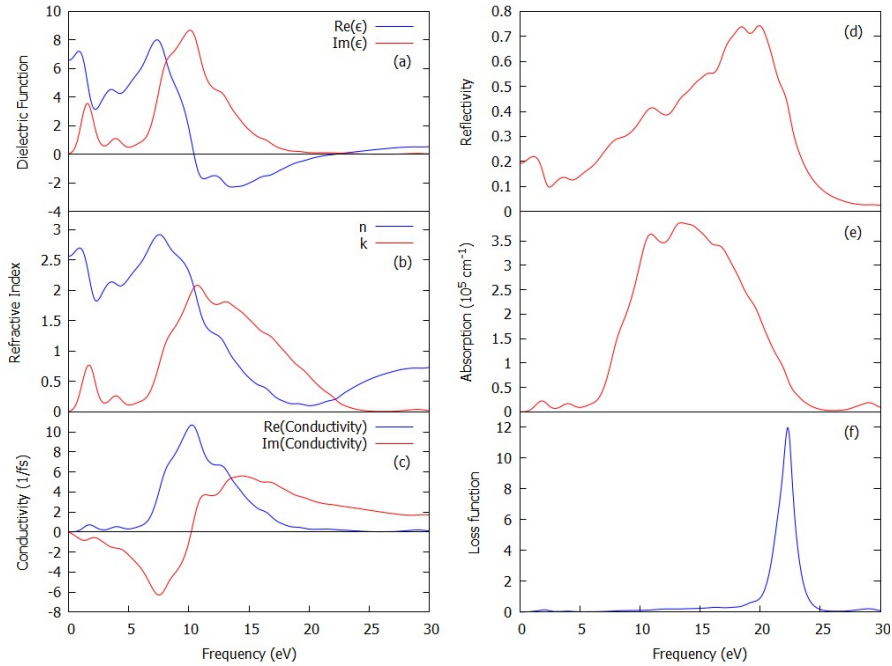

**FIG. 5.** The frequency dependent (a) dielectric function (real & imaginary parts), (b) refractive index (real & imaginary parts), (c) optical conductivity (real & imaginary parts), (d) reflectivity, (e) absorption coefficient, and (f) loss function of $NaC_6$ at 40 GPa.

Figures 5 and 6 illustrate the nature of the optical parameters at two different pressures. From Figure 5, at a pressure of 40 GPa, the dielectric function approaches zero after 22 eV energy, thus this is the plasma frequency of the material. After 22eV, the material becomes nearly transparent since both the reflectivity and absorption coefficient falls sharply at this energy. The real part of the dielectric constant is a measure of polarization. For a certain range of energies from 10 eV to 22 eV, $R(\varepsilon)$ is negative, implying that the polarization in response to incident electromagnetic fields acts to diminish it. The imaginary part, Im($\varepsilon$), gives a measure of loss of incident energy within the system. The real part of the refractive index is the ratio of the speed of light in the medium to that in free space. The imaginary part of the refractive index is known as the extinction coefficient, which is related to the absorption co-efficient. The real part of the refractive index is quite high for the light spectrum spanning infrared to ultraviolet range. After peaking at around 7 eV, it gradually diminishes at the plasma frequency 22 eV, after which it begins to rise briefly again. The conductivity links the current density to the electric field at various frequencies.

The reflectivity curve gradually rises from the infrared and visible light regions, but the value is rather low. However, starting from 10 eV (ultra violet region), it reaches a modest value of about 40% and continues to increase. We see that $NaC_6$ has high reflectivity approaching ~75% in the ultraviolet region encompassing 15 eV to 21 eV. This indicates that the material under consideration could have potential applications as ultraviolet reflector in this spectral band. The absorption coefficient follows the pattern of the extinction parameter as expected. The optical absorption coefficient is high over a wide spectral band (10 eV – 20 eV) in the ultraviolet region. The photon absorption efficacy of $NaC_6$ under high pressure is comparable to some other compounds [54-56] which have potential to be used as efficient ultraviolet absorber.

The loss function $L(\omega)$ shows a very sharp peak at 22 eV, which is the plasmon energy of the material. This is consistent with the fall of absorption and reflectivity functions around the same energy, which leads to the conclusion that the material is effectively transparent to the incident photons at and above the plasma frequencies.

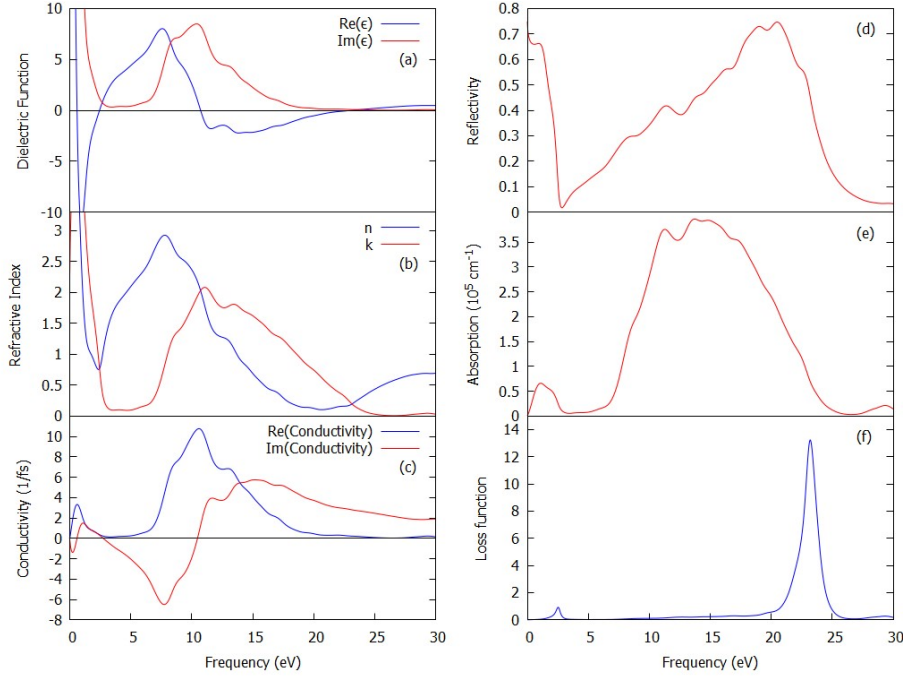

**FIG. 6.** The frequency dependent (a) dielectric function (real & imaginary parts), (b) refractive index (real & imaginary parts), (c) optical conductivity (real & imaginary parts), (d) reflectivity, (e) absorption coefficient, and (f) of $NaC_6$ at 60 GPa.

Increasing pressure gradually shifts the peak of the optical functions towards higher frequencies. Most of these optical properties retain the same general behavior for different frequencies under increasing pressure. However, starting from 50 GPa, the dielectric function, refractive index and the reflectivity undergoes some change, as seen in Figure 6 (showing optical parameters spectra for a pressure of 60 GPa). At 60 GPa, the optical parameters show sharper peaks at low energies in the dielectric constant, refractive index, and optical conductivity spectra. The dielectric function becomes high in the infrared range and decreases sharply towards the end of the visible light spectrum. After that, the dielectric function rises and displays the same general response as it did under lower pressures, and then flattens towards zero near the plasma frequency of approximately 23 eV. It reaches a peak value within 5-10 eV photon frequency. The same pattern is seen in that of refractive index and reflectivity, where they gain large peaks at low frequencies. Interestingly, the reflectivity is now high for visible light in addition to the 10 eV to 20 eV ultraviolet region with a sharp dip at ~4 eV in between. Further increase in pressure does not change the spectral features significantly other than small shifts in the peaks to higher energies including the peak in the loss function.

*3.6. Pressure effects on superconductivity*

Undoubtedly, the most significant property of $NaC_6$ is its predicted high superconducting critical temperature. The critical temperature of a strongly-coupled superconductor can be calculated using the widely employed McMillan formula [57]:

$$T_c = \frac{\theta_D}{1.45} \exp\left\{-\frac{1.04(1+\lambda)}{\lambda - \mu^*(1+0.62\lambda)}\right\} \qquad (6)$$

where $\theta_D$ is the Debye temperature, $\lambda$ is the electron phonon coupling constant and $\mu^*$ is the repulsive Coulomb pseudopotential. The values of these parameters and the critical temperature are enlisted in Table 5 together with the values obtained in Ref. [12]. Since CASTEP is unable to determine the Eliashberg spectral function [45], we have used the electron phonon coupling constants reported in a previous work [12] to determine $T_c$. The repulsive Coulomb pseudopotential has been calculated using the TDOS values at the Fermi level $N(E_F)$ obtained in this study, as follows [50]:

$$\mu^* = \frac{0.26 N(E_F)}{1 + N(E_F)} \qquad (7)$$

**TABLE 5:** Debye temperature $\theta_D$ in K, electron phonon coupling constant $\lambda$, density of states at Fermi Level $N(E_F)$ in states/eV, Coulomb pseudopotential $\mu^*$, and critical temperature $T_c$ in K at various pressures (GPa) for $NaC_6$.

| Pressure (P) | $\theta_D$ | $\lambda$ [12]• | $N(E_F)$ | $\mu^*$ | $T_c$ | $T_c$ [12]• |
|---|---|---|---|---|---|---|
| 40 | 1010.55 | 1.68 | 2.68 | 0.189 | 80.76 | 102.00 |
| 50 | 1316.05 | 1.58 | 2.75 | 0.190 | 97.44 | 101.00 |
| 60 | 1443.34 | 1.53 | 2.81 | 0.191 | 102.25 | 97.00 |
| 70 | 1500.88 | 1.50 | 2.83 | 0.192 | 103.50 | 95.00 |

•Some of these values are interpolated from the $T_c(P)$ and $\lambda(P)$ curves given in Ref. [12].

We can see from Table 5 that the calculated values of $T_c$ at different pressures agree reasonably well with those found in Ref. [12]. Notable deviation is found for the values at 40 GPa, which we believe is due to the use of low value of $\mu^*$ in earlier works [12,13]. As mentioned in a previous section, repulsive Coulomb interaction reduces $T_c$ since it reduces the effective electron-phonon interaction essential for the formation of phase coherent Cooper pairs. In earlier studies $\mu^*$ was taken ad-hoc to be in the range 0.10 – 0.13. If we take $\mu^* = 0.13$ and keep all other parameters unchanged the predicted $T_c$ turns out to be 97.16 K very close to the value (102 K) obtained by Lu et al. [12].

There are several interesting features in the predicted $T_c(P)$ which needs elaboration. We have found that $N(E_F)$ increases gradually with increasing pressure. For conventional superconductors, the electron-phonon coupling constant $\lambda = N(E_F) V_{\text{e-ph}}$, where $V_{\text{e-ph}}$ is the electron-phonon interaction energy [45]. On the other hand, $\lambda$ decreases with increasing pressure [12,13]. This implies that even though $N(E_F)$ increases with pressure, $V_{\text{e-ph}}$ decreases at a greater rate with increasing pressure, the overall effect is a pressure induced decrement of $\lambda$. This decrement can, at least partly, be explained considering the anharmonicity and softening of phonon modes in $NaC_6$. It is observed in Table 4 that the Grüneisen parameter ($\gamma$) is high at 40 GPa and decreases with increasing pressure suggesting that lattice anharmonicity and soft-phonon modes are diminished as pressure increases. This is also supported by the increase in the Debye temperature with increasing pressure (Table 4). There is significant evidence that the electron-phonon coupling constant is enhanced in the presence of soft-phonon modes [58]. This observation follows from the pioneering work by Hopfield [59]. In terms of the Hopfield parameter, $\eta$, the electron-phonon coupling constant can be expressed as, $\lambda \sim \eta/<\theta^2>$ [59]; $<\theta^2>$ is the average squared phonon frequency expressed in temperature showing that presence of soft-phonon modes raises $\lambda$. Furthermore, the Debye temperature increases significantly with pressure but the effect of this increment is insignificant on the predicted values of $T_c$. This is indicative that the positive contribution of $\theta_D$(P) to $T_c$(P) is counterbalanced by the negative contribution due to deceasing $\lambda$(P) and increasing $\mu^*$(P) as pressure is increased. Similar arguments have been put forward to explain the predicted $T_c$(P) behavior in recent studies [12,13].

## 4. Conclusions

On the basis of DFT based first-principles calculations we have carried out an investigation of the structural, elastic, electronic, optical, thermophysical and superconducting state properties of binary $NaC_6$ compound in the sodalite-like structure over a wide range of pressure. The compound is found to be elastically stable only for pressures at and above 40 GPa. In the stable phase $NaC_6$ possesses significant bonding anisotropy. The compound is predicted to be highly ductile and machinable with dominant ionic/metallic bondings. The electronic band structure reveals metallic features in agreement with previous studies [12,13]. The electronic density of states at the Fermi level and the Coulomb pseudopotential are found to increase with increasing pressure. The Debye temperature also increases with pressure. The Grüneisen parameter, on the other hand, decreases with increasing pressure in the elastically stable phase. $NaC_6$ is predicted to be an efficient absorber of electromagnetic radiation in the mid-ultraviolet region. At high pressures, the compound is also a good reflector of visible light. The superconducting transition temperature has been computed at different pressures and discussed in terms of the pressure induced variations in $N(E_F)$, $\lambda$, $\theta_D$, and $\gamma$.

We hope that this study will inspire the researchers to study similar carbon-rich compounds in the sodalite-like structure in greater details in near future.

## Acknowledgements

N. S. K. acknowledges the Research Fellowship from Semiconductor Technology Research Centre, University of Dhaka, Bangladesh, S. H. N. and R. S. I. acknowledge the research grant (1151/5/52/RU/Science-07/19-20) from the Faculty of Science, University of Rajshahi, Bangladesh, which partly supported this work.

## Data availability

The data sets generated and/or analyzed in this study are available from the corresponding author on reasonable request.

**Author Contributions**

M. S. K. performed the theoretical calculations, contributed to the analysis and draft manuscript writing. B. R. R. performed the theoretical calculations, contributed to the analysis, and contributed to manuscript writing. I. M. S. supervised the project and contributed to finalizing the manuscript. R. S. I. contributed to analysis and manuscript writing. S. H. N. designed and supervised the project, analyzed the results and finalized the manuscript. All the authors reviewed the manuscript.

**Competing Interests**

The authors declare no competing interests.